\documentclass[prd,showpacs,nofootinbib,superscriptaddress,twocolumn,preprintnumbers]{revtex4}
\usepackage[mathscr]{eucal}
\usepackage[latin1]{inputenc}
\usepackage{amsmath}
\usepackage{amsfonts}
\usepackage{amssymb}
\usepackage{pictex}
\usepackage{graphicx}

\def\b{{\beta}}
\def\be{\nopagebreak[3]\begin{equation}}
\def\ee{\end{equation}}
\def\ba{\nopagebreak[3]\begin{eqnarray}}
\def\ea{\end{eqnarray}}

\def\d{{\rm d}}

\newcommand{\teta}{\rlap{\lower2ex\hbox{$\,\tilde{}$}}\eta{}}

\def\Tr{{\rm Tr\,}}

\def\g{\gamma}
\def\lp{{\ell}_{\rm Pl}}
\def\q{\mathring{q}}
\def\e{\mathring{e}}
\def\ow{\mathring{\omega}}

\newcommand{\rcr}{\rho_{\mathrm{crit}}}

\newcommand{\f}{\frac}

\usepackage{enumerate}
\DeclareMathOperator*{\sgn}{sgn}

\usepackage{colordvi}
\newcounter{mnotecount}[section]

\newcommand{\comment}[1]{}
\def\f{\frac}

\def\epsilon{\varepsilon}

\begin{document}
\preprint{\vbox{\baselineskip=12pt \rightline{IGC-08/2-1}
\rightline{PI-QG-74}
}}

\title{Is loop quantization in cosmology unique?}
\author{Alejandro Corichi}\email{corichi@matmor.unam.mx}
\affiliation{Instituto de Matem\'aticas,
Unidad Morelia, Universidad Nacional Aut\'onoma de
M\'exico, UNAM-Campus Morelia, A. Postal 61-3, Morelia, Michoac\'an 58090,
Mexico}
\affiliation{Center for Fundamental Theory, Institute for Gravitation and the Cosmos,
Pennsylvania State University, University Park
PA 16802, USA}
\author{Parampreet Singh}
\email{psingh@perimeterinstitute.ca}
\affiliation{Perimeter Institute for
Theoretical Physics, 31 Caroline Street North, Waterloo, Ontario
N2L 2Y5, Canada}

\begin{abstract}
We re-examine the process of loop quantization for flat isotropic models in cosmology.
In particular, we contrast different inequivalent `loop quantizations' of these simple models through their respective successes and limitations and assess
whether they can lead to any viable physical description.
We propose three simple requirements which any such admissible quantum model should satisfy:
i) independence from any auxiliary structure, such as
a fiducial interval/cell introduced to define the phase space when integrating over
non-compact manifolds;
ii) existence of a well defined classical limit and
iii) provide a sensible ``Planck scale" where quantum gravitational effects become manifest.
We show that even when it may seem that one can have several
possible loop quantizations, these physical
requirements considerably narrow down the consistent choices.
Apart for the so called improved dynamics of LQC,
none of the other available inequivalent loop quantizations pass above tests, showing the limitations of lattice refinement models to approximate the homogeneous sector and
loop modified quantum geometrodynamics. We conclude
that amongst a large class of loop quantizations in isotropic cosmology,  there is a unique consistent choice.
\end{abstract}

\pacs{04.60.Pp, 04.60.Ds, 04.60.Nc 11.10.Gh.}
\maketitle

\section{Introduction}

In recent years, loop quantum gravity (LQG) has risen as a
candidate for describing the quantum degrees of freedom of the
gravitational field \cite{lqg}. A theory motivated by LQG
and incorporating several of its physical principles
has been successfully brought to completion in the context  
of homogeneous and isotropic spacetimes coupled to a
massless scalar field (with and without the cosmological constant) \cite{aps0,aps,aps2}. This
approach known as {\it Loop
Quantum Cosmology} (LQC) is based on implementing the methods of LQG in symmetry reduced models \cite{lqc,abl}.
The starting point for
LQC (and LQG) is to express the classical phase space in terms of Ashtekar variables -- the connection and the triad -- and use
holonomies of the connection and fluxes of the triad as the elementary variables for quantization.
The resulting quantum theory (sometimes referred to as polymer representation) is inequivalent to the standard Wheeler-DeWitt quantization
(Fock representation)  even at the kinematical level \cite{AFW,CVZ-2}.
However, the relation between LQC
and full LQG is still an open question, given that to date, there is no
precise way of deriving the homogeneous theory from the full
theory (however, see \cite{engle,tim, brune} for recent attempts).

One of the most dramatic result in LQC (coupled to a massless scalar
field), is  the resolution of the singularity and the
existence of a generic bounce for most states of the theory when the spacetime curvature becomes close to Planck
\cite{aps2,slqc}.
The underlying quantum constraint is a difference equation in the ``geometric'' representation (and
a differential equation in ``connection'' representation).
The resulting physics in LQC is tied to the properties of the difference equation and hence the discretization originating from the quantum theory.
Here the issue of the underlying ambiguities of the
quantum theory becomes important.
It is well known that there exist ambiguities in the definition of the Hamiltonian constraint in LQG (see for instance \cite{ambi}), which is a consequence of dealing with a field theoretic description.
For instance one expects to have both UV and IR ambiguities in the definition of the theory. A natural expectation is that for simple cases,  such as homogeneous cosmologies, one may be able to
achieve a  control over the ambiguities left from the reduction of the problem.
This is, from our perspective, an important motivation to
systematically study loop quantization methods for cosmological models and
address some of these issues.
Even when the minisuperspace approximation reduces much of the freedom
present in a field theory, there are still many inequivalent prescriptions.

In the most simple examples of loop quantization of cosmological models, it has been found that, in the resulting theory {\it after} defining the Hamiltonian constraint,
a geometrical variable (such as volume or area) becomes discretized in uniform steps.
For instance,
in the new or  `improved quantization' of LQC \cite{aps2}, the resulting quantum constraint is uniformly discrete in volume, whereas in the original quantization of LQC it was uniformly discretized in area.
It is then relevant to further study
these inequivalent quantizations and have control over them. We should note that
this ambiguities are over and above the standard factor ordering ambiguities that
will always be present, but that do not change qualitatively the underlying physics.
In what follows we will not concentrate our attention in factor ordering issues.

An important step to gain control over these ambiguities 
is the possibility of mimicking them, 
within the reduced theory, by a suitable but otherwise arbitrary 
choice of variables, followed by a `loop quantization' adapted to them.
Such a construction involves two steps. 
The first one is the identification of a `canonical pair'
of coordinates on phase space and the second step can be described as a `polymerization' of one of this variables \cite{CVZ-2}. For example, one could take the original Wheeler-DeWitt variables, or some other canonical pair and arrive, via this prescription,
to inequivalent quantizations. Is there
a particular choice of coordinates that is selected from both mathematical and physical considerations?
Can any other such arbitrary discretization/quantization of the reduced model be physically viable?
Can we draw general criteria guiding us to obtain a physically consistent quantization of symmetric models? What lessons can we take over to the full theory?

To address these questions  is the main purpose of this manuscript. We
shall argue in detail, employing the well understood case of isotropic
cosmology, that an implementation of physically motivated
considerations in the definition of the Hamiltonian constraint, leads to physically
sensible results. It is simply not true that there exists
a large choice of possibilities that yield mathematical {\it and}
physically consistent results (for the most symmetric cases),
as is sometimes expected and stated in the literature.
To be precise, for the isotropic models various inequivalent quantizations
have been proposed, each of them resulting in a uniform discretization in a different geometric
variable. These include the original quantization of LQC (sometimes referred to as `$\mu_0$' or the old quantization \cite{abl,aps}), the improved quantization of LQC (also known as `$\bar\mu$'  or the new quantization \cite{aps2}), the
quantization obtained by a loop inspired discretization of scale factor in the Wheeler-DeWitt
quantum cosmology \cite{hw,CVZ-2} and
the  lattice refinement models, motivated by LQG,  which allow a large class of  discretizations \cite{lattice}. Here we prove that for a large one parameter family of possible
``quantization prescriptions", that includes all the above cases,
there is {\it only one choice} that satisfies, simultaneously,
three natural
requirements. This is the improved quantization of LQC \cite{aps2}.

The conditions that we put forward as a criteria for
the physical viability of a theory are rather natural and simple to state. We shall
ask that the resulting model be such:\\
i) That the predictions about physical entities which do not depend on `auxiliary structures' should be independent, in the quantum theory,
of any choice related to such structures. The `auxiliary structures' include the  fiducial interval/cell introduced to construct phase space variables by integrations over a non-compact manifold. Examples of physical entities independent of fiducial cell/interval are the spacetime curvature and the energy density. Hence any prediction associated with them, such as the spacetime curvature scale or the energy density at which bounce occurs in a loop or loop inspired quantum cosmological model shall be independent of
the choice of the fiducial cell. This can be seen as a criteria for mathematical consistency of the model.\\
ii) That the quantization prescription gives a well defined
notion of `Planck scale', that is, the scale for which `quantum
gravitational corrections' should become important; and\\
iii) That there exists a well defined classical limit that approximates general
relativity when spacetime curvatures are small.

The last two requirements are motivated by physical considerations that the theory
be well behaved in both UV (planck scale) and IR (general relativity) regimes.
As we will show, these conditions are very stringent requirements to be met by any quantum model. As stated before
these are sufficient to single out a unique quantum theory, from a one parameter family, in the isotropic models.
Some of these consistency conditions we put forward were already used to find the improved quantization
of isotropic LQC from the failures of the old quantization \cite{aps2}. Here these conditions are carried over further
as guiding principles to seek answers to the various questions raised above. As we will show all known loop or `loop inspired' quantizations of isotropic models except the improved quantization  fail to provide a
consistent description {\it and} give physically ill-defined
results.

In order to achieve this, we will employ two available tools that have proved useful when
analyzing isotropic models in LQC. In order to test its mathematical consistency,
the first one is a direct analysis of the resulting
quantum theory respecting its dependence on auxiliary structure. The second one is the use
of {\it effective equations}. These are ``classical'' equations of motion
derived from an {\it effective Hamiltonian constraint},
using the standard Hamiltonian formulation,
that
approximate the dynamical evolution of semiclassical states of the quantum theory. In the
isotropic models, using the numerical simulations of the semi-classical states at late times, effective Friedman equations are shown to approximate the
quantum dynamics to an exceptional success and lead to classical Friedman
dynamics at low curvatures \cite{aps,aps2,vp}
(for analytical derivations of such equations from quantum theory see \cite{vt}).
In what follows we make extensive use of this correspondence in order to
analyse the properties of the effective dynamics and reach conclusions about the full quantum dynamics.
In particular, this will allow us to test the viability of different quantizations in both the UV, Planck regime and in the IR, general relativity limit. Further, it provides viability tests even in the case where the use of non-compact cell is not necessary or one chooses a closed topology.

The lessons from the isotropic model
when taken at face value open the prospects of application to the
construction of less symmetric models, such as anisotropic cosmologies
and the interior of black holes. One could also hope that these guiding principles will be useful to select the physically viable theory from the many possibilities available for the full theory.

The structure of the paper is as follows. To make this manuscript self-contained,
in Sec.~\ref{sec:2} we provide a pedagogical review of the quantization of $k=0$ FRW cosmology\cite{aps2}.
Sec.~\ref{sec:3} is devoted to the
study of the underlying freedoms of the auxiliary structures and their effects on the physics
of various
quantizations. We show that different choices of basic variables used in the inequivalent quantizations have
different transformation properties, and that there is a unique choice (within the considered
set of variables) for which the resulting predictions are physically. This unique choice of variables is the one forced on us by the improved quantization of LQC. In Sec.~\ref{sec:4} we consider, from the perspective of the Friedman dynamics, expected to
be valid in the low curvature sector of the theories, the way different choices of basic
variables affect the
dynamics at low spacetime curvatures.
This shows that all choices, except the one used in improved dynamics, are
simply untenable if one wants to recover the low curvature, GR limit for matter satisfying
null energy condition. We summarize the results and highlight open issues in construction of anisotropic and black hole interior models in Sec.~\ref{sec:7}.

\section{Isotropic and Flat Model: A brief review}
\label{sec:2}

We consider a $k$ = 0 homogeneous and isotropic FRW cosmological
model with a 3-manifold $\Sigma$ which is topologically $R^3$. In
order to define the symplectic structure it is necessary to fix a
fiducial cell ${\cal V}$. We can introduce a flat fiducial metric
$\q_{ab}$ on the manifold with respect to which the coordinate
volume of ${\cal V}$ is $V_o = \int_{\cal V}\, \sqrt{\q}\, \d^3\! x$. The FRW
spacetime is described by the metric
\be
\d s^2 = - N^2 \d t^2 + a(t)^2 \d {\bf x}^2
\ee
where $N$ is the lapse function and $a$ is
the scale factor of the universe. Note that in
the action framework, choice of the fiducial cell amounts to choosing
the limits of integration in the integral over spatial coordinates:
\be \label{action} S
= \nonumber \f{1}{16 \pi G} \, \int \d t\, \d^3 x \sqrt{|g|}\, R =
\nonumber \f{V_o}{16 \pi G} \, \int \d t\, N\, a^3\,  R ~. \ee

The gravitational part of the conventional Wheeler-DeWitt phase space
consists of $a$ and its
conjugate $P_a = - 3 V_o \, a \,\dot a/(4 \pi G N)$.
In order to write the phase space in terms of the Ashtekar-Barbero
variables we first introduce a fiducial triad $\e^a_i$ and co-triad
$\ow^i_a$ compatible with $\q_{ab}$. The conjugate phase space variables are the
$SU(2)$ connection $A^i_a = \Gamma^i_a + \gamma K^i_a$ and the
densitized triad $E^a_i$ satisfying
\be
\{ A^i_a(x),E^b_j(y)\} = 8
\pi G\, \gamma\, \delta^b_a \delta^i_j \delta^3(x,y) ~.
\ee
Here $\Gamma^i_a$ is the spin connection measuring the intrinsic
curvature (which vanishes in the $k=0$ model), $\gamma$ is the
Barbero-Immirzi parameter and $K^i_a$ is the extrinsic curvature
1-form related to extrinsic curvature $K_{ab}$ as $K^i_a = e^{b i}
K_{ab}$ where $e^a_i$ is the un-densitized triad.

Due to the underlying symmetries of the manifold, the connection and the triad
can be written as \cite{abl}
\be\label{AE_defs}
A^i_a \, = \, \tilde c  \, \ow^i_a, ~~~~
E^a_i \, = \, \tilde p \, \sqrt{\q} \, \e^a_i ~.
\ee
It is convenient to introduce
\be
c = V_o^{1/3} \, \tilde c \, ~~ \mathrm{and} ~~ p = V_o^{2/3} \, \tilde p ~.
\ee

In order to find the relationship between the triad components and
the scale factor it is useful to note that \be E^a_i E^{b i} = q
\, q^{a b} \ee which on using (\ref{AE_defs}) implies
\be\label{pa2} |p| = V_o^{2/3} \, a^2 ~. \ee

Similarly, one can compute the extrinsic curvature which turns out
to be \be K_{a b} = \f{a \, \dot a}{N} \ow_a^i \ow_{b i} \ee
leading to $K^i_a =  (\dot a/N) \ow_a^i$. Using (\ref{AE_defs}) the
connection component gets related to the rate of change of scale
factor as \be\label{cdota} c = \gamma \, V_o^{1/3} \f{\dot a}{N}
~, \ee
holding only for the physical solutions of general relativity (GR).
The phase space is  characterized by the conjugate  variables $c$
and $p$ satisfying:
\be
\{c,p\} = \f{8 \pi G \gamma}{3} ~.
\ee

Following the route to quantization as in LQG, an important feature which
emerges is that there exists no $\hat c$ operator but only the holonomies
of the gravitational connection
\be
h_k^{(\lambda_c)} = \cos (\lambda_c \, c/2) \mathbb{I} + 2
\, \sin  (\lambda_c \, c/2) \tau_k
\ee
where the holonomy\footnote{Since
we will soon study various choices of phase
space variables and resulting quantizations, to keep a track of
different $\lambda$'s in holonomies a subscript on $\lambda$ is introduced.} is computed along the edge $\lambda_c \e^a_k$
and $\tau_k = - i \sigma_k/2$, where $\sigma_i$ are the Pauli spin
matrices.
These generate an algebra of the almost periodic functions whose
elements are of the form $\exp(i \lambda_c \, c/2)$. The resulting
kinematical Hilbert space is $L^2({\mathbb R}_{\mathrm{Bohr}},\d
\mu_{\mathrm{Bohr}})$, a space of square integrable
functions on the Bohr compactification of the real line. In this
space the eigenstates of $\hat p$ labeled by $|\mu\rangle$
satisfy $\langle \mu_1|\mu_2 \rangle = \delta_{\mu_1,\mu_2}$. The
action of fundamental operators on the eigenstates $|\mu\rangle$ is
\be
\label{p_act} \hat p| \mu \rangle = \f{8 \pi \gamma \lp^2}{6}
\mu |\mu \rangle ~, ~~
\ee
\be\label{exp_act}
{\widehat{\exp(i \lambda_c \, c/2)}}
|\mu\rangle = |\mu + \lambda_c \rangle
\ee
and
\ba \widehat h_k^{({\lambda}_c)} |\mu\rangle &=& \nonumber
\f{1}{2} \left(|\mu + \lambda_c\rangle + |\mu - \lambda_c\rangle
\right) \mathbb{I}
\\&& + \f{1}{i}  \left(|\mu + \lambda_c\rangle - |\mu -
\lambda_c\rangle \right) \tau_k ~. \ea In order to obtain the
quantum constraint the key step is to rewrite the classical
gravitational constraint with field strength $F_{ab}^i$ of the connection
\be \label{eq:cgrav}
C_{\mathrm{grav}} = - \gamma^{-2} \int_{\cal
V} \d^3 x\,  \epsilon_{ijk} \,\f{E^{ai}E^{bj}}{\sqrt{|\det E|}}\,
F_{ab}^i ~ \ee in terms of holonomies and triads and then
quantize (where we have chosen $N=1$). The matter part of the constraint is quantized in a
similar way. The gravitational constraint is composed of two
terms. The term involving inverse triad can be rewritten as \ba
\label{cotriad} \epsilon_{ijk}\, \,\f{E^{aj}E^{bk}}{\sqrt{|\det
E|}}\, &=& \nonumber \sum_k \f{({\rm sgn}\,p)}{2\pi\gamma
G\lambda_c\, V_o^{\f{1}{3}}}\, \, \mathring\epsilon^{\,abc}\,\,
\ow^k_c\, \times \\ \, &&  \, \Tr\left(h_k^{(\lambda_c)}\,
\{h_k^{(\lambda_c)}{}^{-1}, V\}\, \tau_i\right) \ea using an
identity on the classical phase space. The field
strength can be classically written in terms of a trace of
holonomies over a square loop $\Box_{ij}$, considered over a face
of the elementary cell, with its area shrinking to zero: \be
\label{F} F_{ab}^k\, = \, -2\,\lim_{Ar_\Box
 \rightarrow 0} \,\, \Tr\,
\left(\f{h^{(\lambda_c)}_{\Box_{ij}}-1 }{\lambda_c^2 V_o^{2/3}} \right)
\,\, \tau^k\, \ow^i_a\,\, \ow^j_b\,  \ee
with
\be
h^{(\lambda_c)}_{\Box_{ij}}=h_i^{(\lambda_c)} h_j^{(\lambda_c)}
(h_i^{(\lambda_c)})^{-1} (h_j^{(\lambda_c)})^{-1}\, . \ee
The classical gravitational constraint thus becomes

\be \label{hc1} C_{\mathrm{grav}} = \nonumber \lim_{Ar_\square
\rightarrow 0} C_{\mathrm{grav}}^{(\lambda_c)} \ee
with
\ba\label{cgc}
C_{\mathrm{grav}}^{(\lambda_c)}
&=& \nonumber  \sin (\lambda_c c) \Bigg[-\, \f{1}{2\pi G\g^3}
\f{{\rm sgn}(p)}{\lambda_c^3}\, \times \\ && \sum_k \Tr \tau_k
h_k^{(\lambda_c)}\, \{(h_k^{(\lambda_c)})^{-1},\, V\}\Bigg] \sin
(\lambda_c c) ~.
\ea
Since the underlying geometry in the quantum theory resulting
from LQG is discrete, the loop $\Box_{ij}$ can be
shrunk at most to the area which is given by the minimum eigenvalue of the area operator in LQG:
$\Delta = \tilde\kappa\, \lp^2$ with $\tilde\kappa$ of order one.\footnote{It has been standard in the LQC literature to choose $\tilde\kappa= 2 \sqrt{3} \pi \gamma$  \cite{abl}, but
it can also be taken as a parameter to be determined \cite{slqc}.}
The area of the loop with
respect to the physical metric is $\lambda_c^2 |p|$. Requiring
the classical area of the loop $\Box_{ij}$ to have the quantum area gap
as given by LQG, we are led to set
$\lambda_c = \sqrt{\Delta/|p|}$. Since $\lambda_c$ is now a
function of triad, the action of $\exp(i \lambda_c(p) c)$ becomes
complicated on the states in triad ($\mu$) basis. However, its
action in volume ($\nu$) basis is very simple: it drags the state
by a unit affine parameter.

It is then convenient to introduce
\be
\b := \f{c}{|p|^{1/2}}
\ee
such that $\lambda_c c = \lambda_\b \b$ where $\lambda_\b := \sqrt{\Delta}$ is the
new affine parameter\footnote{Note that in \cite{slqc}, where
it was first introduced, the symbol
{\tt b} was used to denote the object $\b$. We shall from now on employ the new
notation.}. Note that $\b$ is  conjugate variable to $\nu$,
satisfying  $\hbar \{\b,\nu\} = 2$, where $\nu$ labels the eigenstates of
the volume operator
\be
\hat V \, |\nu\rangle = 2 \pi \lp^2 \gamma |\nu| \, |\nu\rangle ~.
\ee
The action of the exponential operator then becomes very simple:
\be
\widehat{\exp(i \lambda_c c/2)} \, |\nu\rangle ~ = ~\widehat{\exp(i \lambda_\b \b/2)}
\, |\nu\rangle ~ = ~ |\nu +  \lambda_\b\rangle ~.
\ee

Further, all of the identities used to write classical constraint
in terms of holonomies remain unaffected and the gravitational
quantum constraint operator with the following action is obtained
\be \label{qgc}
\hat C_{\mathrm{grav}} \Psi(\nu,\phi) = \sin(\lambda_\b \b)
A(\nu) \sin(\lambda_b \b) \Psi(\nu,\phi)\ee where $\phi$ refers to a matter
field and
\be
A(\nu) = - \f{6 \pi \lp^2}{\gamma \lambda_b^3}\,
|\nu| ~ \left||\nu + \lambda_\b| - |\nu - \lambda_\b|\right| ~.
\ee
The quantum constraint results in a quantum difference equation
with uniform steps in $\nu$:
\be C^+(\nu) \Psi(\nu + 4 \lambda_\beta) +
C^0 \Psi(\nu) + C^- \Psi(\nu - 4 \lambda_\beta) = \hat C_{\mathrm{matt}}
\, \Psi(\nu) \ee
where $C^{\pm}$ and $C^0$ are functions of
$|\nu|$ \cite{aps2} \footnote{The analysis in Ref. \cite{aps2} was
performed in a slightly different convention. To compare $v$ in
\cite{aps2} is related to $\nu$ as $v = 4/\sqrt{8 \pi \gamma
{\sqrt 3}} \nu/\lp$.}. The resulting quantization was first
introduced in Ref. \cite{aps2} and redressed in $(\b,\nu)$
variables recently \cite{slqc}. It is often referred to as the
improved or the new quantization of LQC\footnote{Due to choice of
conventions in Ref. \cite{aps2}, it is also known as $\bar \mu$
quantization.} \cite{aps2}. It has been extensively studied for
the case of a massless scalar field with and without the
cosmological constant \cite{aps2,polish,vp} and has also been extended
to the case of a closed \cite{closed} and open models \cite{open} and the massive scalar
field \cite{massive}. In this quantization (and in the old quantization \cite{aps})
it is possible to define a notion of time, an inner product, Dirac
observables and and study the states in the physical Hilbert space. One can thus extract
physical predictions  from the theory.

We highlight the main features of this quantization for the massless scalar
$\phi$ whose Hamiltonian is
\be
H_{\mathrm{matt}} = \f{C_{\mathrm{matt}}}{16 \pi G} = \f{P_\phi^2}{2 |p|^{3/2}} ~
\ee
with
\be\label{p_phi}
P_\phi = V_o a^3 \dot \phi.
\ee
Here $V_o$ appears by integration over the spatial coordinates in the action integral (as in the
gravitational case).

\begin{enumerate}
\item For states which are semi-classical at late times, i.e. those
which lead to a large classical universe like ours, the backward
numerical evolution via the quantum difference equation leads to a
quantum bounce when energy density of the field becomes equal to
$\rcr \approx 0.82 \rho_{\mathrm{Pl}}$.

\item When curvatures become much smaller than the Planck curvature
(or for $\rho \ll \rcr$)
the expectation values of the Dirac observables agree with the
values obtained from classical GR.

\item The role of modifications pertaining from inverse triads in
the gravitational and matter part of the constraint is totally
suppressed in comparison to those originating from the field
strength. In fact, even if one does not use (\ref{cotriad}) to
rewrite inverse triad operators one finds negligible change in the
behavior of expectation values for above states.\footnote{The
underlying reason is that corrections coming from field strength
are significant when curvature or energy density becomes of the
order Planck. Contrary to this, corrections coming from inverse
triad are not tied to any curvature scale in the flat model. Only
when there exists an intrinsic curvature as in closed model these
corrections are meaningful and can lead to potentially interesting phenomenological
effects \cite{jm_ps}.}

\item Using the previous observation, one can write an exactly solvable model of LQC
which is based on the mild approximation of ignoring modifications
coming from the inverse triad. This model known as Simplified LQC
(sLQC) descends directly from LQC after
an approximation is performed at the quantum level \cite{slqc}. Alternatively, one
can arrive to this system by choosing from the outset the lapse function as $N=|p|^{1/2}$, for which Hamilton's equations describe evolution with respect to the scalar field $\phi$ \cite{AA:pc}.
Analysis of this model shows:\\

(a) The bounce is not restricted to semi-classical states but occurs for
states in a dense sub-space of the physical Hilbert space. \\

(b) There exists a supremum of the expectation value for the
energy density. This supremum $\rho_{\mathrm{sup}} = \sqrt{3}/(16
\pi^2 \gamma^3 G^2 \hbar) = \rcr$ (using  standard values of minimum eigenvalues of
the area gap which fixes $\lambda_\b$, the only free parameter of the model). 
We note that existence of an
absolute maximum of the energy density in this cosmological model
implies non-singular evolution, in terms of physical quantities.\\

(c) States that evolve to be semiclassical at late times, as determined by the
dispersion in canonically conjugate observables, have to evolve from
states that also had semiclassical properties  before the bounce (even when there might be  
asymmetry in their relative fluctuations without affecting semiclassicality)
\cite{recall}.\footnote{Using another simplified model which is not obtained from LQC \cite{harmonic}, it is
possible to show, that for a universe as large as 1 MPc if one evolves a
semiclassical state from late times in the expanding branch then the change in the square of
relative dispersion of volume observable before the bounce turns out be
$10^{-113}$. Semiclassicality is preserved to an amazing degree across the
bounce.}

\item Using geometric methods of quantum mechanics one can write an
effective Hamiltonian which provides an excellent approximation to the
behavior of expectation values of Dirac observables in the
numerical simulations \cite{vt}. The effective Hamiltonian is\footnote{The effective
Hamiltonian will in principle also
have contributions from terms depending on the properties of the state such as its
spread. Effect of these terms
turns out to be negligible as displayed from the detailed numerical analysis \cite{aps2,closed}.}
\be \label{effham}
\f{3}{\gamma^2} \, \f{\sin^2(\lambda_\b \b)}{\lambda_\b^2} |p|^{3/2} = 8 \pi G
\, H_{\mathrm{matt}} ~
\ee
which leads to modified Friedman and Raychaudhuri equations on computing the
Hamilton's equations of motion. Using (\ref{effham}) one can find that the
energy density $\rho = H_{\mathrm{matt}}/|p|^{3/2}$ equals
$ 3 \sin^2(\lambda_\b \b)/(8 \pi G \gamma^2 \lambda_\b^2)$. Since the latter has a
maxima equaling $3/(8 \pi G \gamma^2 \lambda_\b^2)$ using $\lambda_\b = \sqrt{\Delta}$
we find that the maximum energy density obtained from effective Hamiltonian is
identical to $\rho_{\mathrm{sup}}$ in sLQC.
\end{enumerate}

This summarizes the main features of the resulting quantizations for
the LQC. Let us now explore various inequivalent quantizations. For that it
is important to understand underlying freedoms in these models.

%In a sense, we can say that the was a UV problem with the old quantization,
%namely that it d and that it also had
%an IR problem, given that it gave quantum effects
%where one would expect classical GR to hold. The new quantization solves, with
%one stroke, both problems.

%{\bf (may be we can say first issue is UV problem and second IR,
%and new quantization solves them together)}

\section{Underlying freedoms and Various parametrizations}
\label{sec:3}

For the classical phase space of isotropic models there are two underlying freedoms. As it will turn out their
resulting implications  play an important role in the physical viability of any given quantization.
This has been stressed before  in comparison between new and old quantizations of LQC \cite{aps2}. Here our aim
will be to generalize to a large class of inequivalent quantizations.

The underlying freedoms are:\\
(i) {\it Freedom of the choice of spatial coordinates:} With this one means that given the metric
\be
g_{ab} = -N^2\, \nabla_a t\nabla_b t + a(t)^2 \, q^o_{ab}
%\d s^2 = - N^2 \, \d t^2 + a(t)^2 \, \d {\bf x}^2
\ee
we have a freedom to rescale the coordinates leaving the metric invariant. This implies
\be\label{x_s}
x ~ \rightarrow ~ x' ~ = ~ l \, x
\ee
with scale factor scaling as
\be\label{a_s}
a ~ \rightarrow ~ a' ~ = ~l^{-1} \, a ~.
\ee
Under this rescaling of the coordinates, the  coordinate volume of the fixed
fiducial cell ${\cal V}$ changes:
\be\label{V_s}
V_o = \int_{\cal V} \,\sqrt{\q}\, \d^3 x \, \rightarrow \, V_o' \, = \, l^3 \, V_o ~.
\ee
However, the physical volume of the fiducial cell is invariant
\be
V \, = \, a^3 \,  V_o \, \rightarrow V' \, = \, a'^3 \, V_o' = a^3 \, V_o = V
\ee

(ii) {\it Freedom of the choice of the fiducial cell:}
Apart from rescaling the coordinates, we can choose
a different cell to define the symplectic structure.
The new fiducial cell can be larger or smaller than ${\cal V}$ (without changing
its cubical shape) which amounts to changing the limits of integration over the spatial
coordinates uniformly in Eq.(\ref{action}).
This freedom implies:
\be
{\cal V} \, \rightarrow \, {\cal V'} \, \, \, ~ {\mathrm{such}} ~ {\mathrm{that}} ~ \, \, \,  
V_o'\, = \,\alpha^3 \, V_o ~.
\ee

The choice of coordinates and the fiducial cell defined over non-compact manifold to perform
integrations are the auxiliary background structures in the framework and the resulting physics
should be independent of their choice.

From (\ref{pa2}) and (\ref{cdota}), and using (\ref{x_s}) and (\ref{a_s})
it is easy to see that under the change of spatial coordinates, $c$ and $p$ are
invariant\footnote{Without any loss of generality, we fix the lapse
to be unity in the following discussion and the next section.}:
\be
c \,  \, %\gamma \, V_o^{1/3} \, \f{\dot a }{N} \, \,
\rightarrow \, c^{\, \prime} \, = \, \gamma \, V_o'^{\, 1/3} \, \dot a'  \,  = \, c
\ee
and
\be
|p| \,  \, %V_o^{2/3} \, a^2  \,
\rightarrow \, |p|' \, = \,  V_o'^{\, 2/3} \, a'^{\, 2}  \, = |p| ~.
\ee
%Thus, $c$ and $p$ are invariant under the change of spatial coordinates.
However, under ${\cal V} \rightarrow {\cal V}'$:
\be\label{c_V_o}
c \, %= \, \gamma \, V_o^{1/3} \, \f{\dot a }{N} \, \,
\rightarrow \, c^{\, \prime} \, = \, \gamma \, V_o'^{\, 1/3} \, {\dot a} \,  = \, \alpha\, c
\ee
and
\be\label{p_V_o}
|p| \, %= \, V_o^{2/3} \, a^2 \, \,
\rightarrow \, \, |p|^{\, \prime} \, = \, V_o'^{\, 2/3} \, a^2 \, = \, \alpha^2 \, |p| ~.
\ee
Similarly, $(\beta,\nu)$ are invariant under the change of coordinates.
The eigenvalue of the volume operator changes proportionally with
the change in the fiducial volume of the cell,
\be
\nu %:= \f{\sgn(\nu) |p|^{3/2}}{2 \pi \lp^2 \gamma}
\rightarrow \nu' =  \f{\sgn(\nu) |p|'^{\, 3/2}}{2 \pi \lp^2 \gamma} = \alpha^3 \, \nu ~.
\ee
However, in contrast to the
behavior of $c$, $\b$ is {\it invariant} under the change of the cell:
\be
\b %:= \f{c}{|p|^{1/2}}
\rightarrow \b' = \f{c'}{|p|'^{\, 1/2}} = \b %\f{\alpha c}{\alpha |p|^{1/2}} = \b
\ee

Let us consider the role of change of cell at the quantum level. As an example, in the new quantization
the parameter $\nu$ provides the physical volume of the cell ${\cal V}$. Under the change ${\cal V} \rightarrow {\cal V}' = \alpha^3 {\cal V}$, the states is still labeled by $|\nu\rangle$ however, the interpretation changes since it gives the volume of the
new cell ${\cal V}'$. In order to relate the quantum theories of cells ${\cal V}$ and ${\cal V}'$ it is possible to define a unitary map ${\cal U}_\alpha$
\be\label{new-map}
{\cal U}_\alpha\,|\nu \rangle := |\alpha^{-3} \nu\rangle
\ee
under which the  operator $\hat \nu' := {\cal U_\alpha}\, \hat{\nu}\, {\cal U_\alpha}^{-1}$ has the following action
\be
{\hat \nu}'\, |\nu\rangle = {\cal U}\, \hat \nu \,|\alpha^3\, \nu\rangle = \alpha^3\, \hat \nu\, |\nu\rangle .
\ee
That is, the operator $\hat \nu$ `scales' just as its classical analog, {\it in the kinematical Hilbert space}.
However, at the level of the physical Hilbert space the mapping in general   mixes the
superselected sectors, and therefore, ceases to exist. One might try to define,
on the superselected sectors a new mapping ${\cal U}_\alpha'$ that leaves the kets invariant
(corresponding to $\alpha = 1$ in (\ref{new-map})), but clearly, with respect to this new map, the operator $\hat{\nu}$ does not scale appropriately.

Elements of the algebra of almost periodic functions, $\exp(i \lambda_\beta \beta)$ are preserved  under
the change of fiducial cell since both $\lambda_\beta$ and $\beta$ are invariant.
This behavior at the quantum level is not unique to $(\beta,\nu)$ variables. It can be checked that even the
operator $\hat p$ scales as its classical analog. The behavior of elements $\exp(i \lambda_c c)$ is however
mores subtle. Under the rescaling of cell, though
the edge labeled by $\lambda_c$ gets modified by $\alpha$ (since by construction it
corresponds to the interval $[0,\lambda_c\,V_o^{1/3}]$ in the corresponding Cartesian coordinate),
the function $c$ rescales in such a way that the new holonomy is associated to
{\it the same function}. Thus, the algebra generated by  $\exp(i \lambda_c c)$ and $p$ is common to all cells.
From the algebraic perspective to quantization, the difference must be in the
states \cite{lqc}, that have a different interpretation in each case.

From the above behavior of $(\beta,\nu)$ and $(c,p)$, it may seem that the resulting quantum  
theories would be very similar, with respect to change of fiducial cell. As we will now show this is not the case. The physics of old quantization of LQC
based on $(c,p)$ turns out to be starkingly different from that of the new quantization based on $(\beta,\nu)$.
For that we have to understand the reason for novel physical effects in the loop quantization, as compared to more standard Schr\"odinger representations. They occur  
when the holonomies used to approximate the field strength `saturate' (and the
approximation fails to be good).
These lead to profound change in physics depending on the underlying variables.

\subsection{Features of the Old Quantization of LQC}
\label{cp_sub}

In the old quantization of LQC, $\lambda_c$ is treated as constant, $\lambda_c = \sqrt{\Delta}$, and plays the role of affine parameter. The holonomies
considered are those of connection $c$  and the  action
of exponential operators is as in (\ref{exp_act}). The
quantum gravitational constraint is of the form (\ref{qgc}) with
operators $\sin(\lambda_c c)$ and in the $p$ representation the
resulting
difference equation has uniform steps in triad eigenvalues. The effective Hamiltonian in this quantization, to the leading order, is
\be \label{effham1}
\f{3}{\gamma^2} \, \f{\sin^2(\lambda_c c)}{\lambda_c^2} |p|^{1/2} = 8 \pi
G \, H_{\mathrm{matt}} ~.
\ee
Though the Hamiltonian looks  similar to (\ref{effham}), the
important difference is that the energy density, $\rho$ now equals $3
\sin^2(\lambda_c c)^2/(8 \pi \gamma^2 \lambda_c^2 G |p|)$. When the holonomies saturate, the latter
depends on the phase space variable. As an example, for the case of massless scalar energy density at
which bounce occurs is $\rcr =
\sqrt{2}(3/(8 \pi G \gamma^2 \lambda_c^2))^{3/2}/P_\phi$ \cite{aps}.
Using $(\ref{p_phi})$ we find that under the change of the fiducial cell:
\be
P_\phi \, \longrightarrow \, P_\phi' \, = \, \alpha^3 \, P_\phi
\ee
which implies that $\rcr$ is not invariant as ${\cal V} \rightarrow {\cal V}'$ and can be changed arbitrarily depending on the cell. This already violates one of our
physical conditions, since the energy density is a quantity that is invariant under the change of cell. The density (and hence the spacetime curvature) at which bounce occurs in this quantization can thus be made as small
as wished by an
appropriate choice of $\alpha$. These unphysical effects are observed in the numerical simulations of the evolution
of states with quantum constraint in the old LQC \cite{aps}.

Dependence of $\rcr$ on phase space variable leads to another problem in the model. The theory does not lead to classical GR at low curvature scales. This is
immediately evident if one considers the case of a positive
cosmological constant. For such a matter source in a spatially
flat FRW universe, classical GR predicts accelerated expansion for
all time in the future. However, the universe fails to expand forever
in the case of old LQC.  It is found that the universe recollapses in the low
curvature regime implying that the theory does not lead to GR. \footnote{This is sometimes incorrectly  attributed in the literature to
the breakdown of mini-superspace approximation which happens when the
universe is very large. However as is very clear from above these unphysical effects
arise due to a flaw in this quantization at a very basic level -- treating $\lambda_c$ as the affine parameter.}
We will elaborate on this limitation in the next Section. Let us now discuss other inequivalent quantizations and compare
them with LQC.

\subsection{Quantization based on  metric variables}

Motivated by the success of LQC, one can ask whether it is possible that
a loop inspired, `polymer quantization' based on $(P_a,a)$, but
such that one rewrites the Wheeler-DeWitt
quantum constraint in terms of the exponential operators
of $P_a$ succeed? Such a quantization has indeed been attempted in the
literature \cite{hw,CVZ-2} and it has been shown that it
leads to a quantum constraint which is
a difference equation with uniform step size in the scale factor with
$\lambda_{P{_a}}$ as the affine parameter.
As we have argued in previous sections, the criteria to check the consistency of such quantization is that the
phase space variables should be invariant under the freedom of rescaling
of spatial coordinates and $P_a$ should be invariant under the freedom
of the choice of fiducial cell. Under the rescaling of coordinates:
\be
a \rightarrow a' = l^{-1} a
\ee
and
\be
P_a  \rightarrow \, P_a' = - \f{3}{4 \pi G N} \,  V_o'\, a'\, \dot a' = - \f{3}{4 \pi G N} \,  l \, V_o\, a\, \dot a = l \, P_a ~.
\ee
Further, under ${\cal V} \rightarrow {\cal V}'$:
\be
P_a \rightarrow P_a' = \alpha^3\, P_a ~.
\ee
Thus, not only is $P_a$  not invariant with respect to the change of the
fiducial cell but the set $(P_a,a)$ is also sensitive to the scaling of
spatial fiducial coordinates. Thus,
if one tries to perform a polymer quantization using $(P_a, a)$ as
basic variables then one is in a deeper trouble than the old quantization of
LQC, since the quantum difference equation one would obtain will change under
the change of spatial coordinates or the fiducial cell.

In any case, if one still proceeds with a loop quantization based on
$(P_a,a)$ variables, one is then led to an effective Hamiltonian:
\be
\f{2 \pi G}{3} \f{\sin^2(\lambda_{P_{a}} P_a)}{\lambda_{P_{a}} a} = H_{\mathrm{matt}} ~.
\ee
Using which it can be shown that the maxima of energy density would occur
at $\rcr = (2 \pi G/3)^3 (2/\lambda_{P_{a}}^6 P_\phi^4)$ if we consider a massless
scalar field. Though both $(c,p)$ and $(P_a,a)$ based quantizations are
inconsistent, since $\rcr$ depends on $P_\phi$ strongly in latter, it shows
that the problems such a the lack of classical limit and a `quantum' bounce at
arbitrarily small densities are even more pronounced in the $(P_a,a)$ quantization.

\subsection{Quantizations based on a lattice refinement model }

There have been attempts in the literature to develop a model of lattice refinements which
takes the viewpoint that the `improved dynamics' of LQC results from  a
special kind of lattice refinement of the original uniform lattice in the triad
variable in the old quantization based on $(c,p)$. The model is inspired by
LQG and the framework {\it assumes} that in the full theory the action of the
Hamiltonian constraint is ``generally'' to create new vertices. In this scheme various
kinds of lattice refinements are allowed and therefore one deals with a generalized set
of phase space variables that `carry' the information of the particular
refinement:
\be
P_g = c \, p^m \,, ~~~~~ g = \f{p^{(1 - m)}}{1 - m}
\ee
obtained from $(c,p)$ by a canonical transformation.\footnote{In this choice
of general variables, we suppress the orientation of the triad following Ref. \cite{lattice}.
The conclusions do not change even if the orientation is taken into account.}
The case $m=-1/2$ corresponds to the action of Hamiltonian which only results in change of the number of vertices without affecting the labels of the edges in the
spin network and the case $m =0$ corresponds to the change only in the edge spins. Since the Hamiltonian acts by a combination of both processes, the constraint on $m$ amounts to $-1/2 < m < 0$ \cite{lattice,darkpatch}.

Let us consider the transformation properties of the variables $(P_g,g)$ for a general $m$. Note that for $m = -1$,
the quantization based on these variables will be equivalent to the one based on $(P_a,a)$.
For any given $m$, under the rescaling
${\cal V} \rightarrow {\cal V}'$ the coordinate $P_g$ transforms as:
\be
P_g \rightarrow P_g' = c' \, p'^{\, m} = \alpha^{2 m + 1} P_g ~.
\ee
Thus, $P_g$ is invariant if only if $m = -1/2$, i.e.  when the variables are equivalent to $(\b,\nu)$. For all other choices of $m$, $P_g$ fails to be invariant.

The classical and quantum constraints with a general set of variables $(P_g,g)$ are of the form (\ref{cgc}) and (\ref{qgc})
with the effective Hamiltonian
\be
\f{3}{8 \pi G} \f{\sin^2(\lambda_{{P_g}} P_g)}{\gamma^2  \lambda^2_{P_{g}}} \, \left((1 - m) g\right)^{(1 - 4 m)/(2(1 - m))} = H_{\mathrm{matt}} ~
\ee
from which one can find the energy density at the bounce. For the massless scalar model it turns out to be\footnote{For another derivation see Ref. \cite{bbq}.}
\be
\rcr = \f{3}{8 \pi G \gamma^2 \lambda^2_{P_{g}}} \, \left(\f{8 \pi G}{6} \gamma^2 \lambda^2_{P_g} \, P_\phi^2\right)^{(2 m + 1)/(2 m - 2)} ~.
\ee
Thus, {\it the critical density at which quantum bounce occurs depends on
$P_\phi$ and hence the fiducial cell unless $m = -1/2$.} Hence ruling out the physical viability of
lattice refined models. We are thus able to prove that
unless the quantization is equivalent to the new quantization of LQC, it suffers from similar problems as we discussed
in the old quantization of LQC.
In a precise sense, all these proposed inequivalent quantizations suffer
from both the ultra-violet problem -- quantum bounce not
at an invariant scale, and also the infra-red problem -- since
they predict quantum gravity effects at arbitrarily low spacetime curvature.\\

{\it Remark:} A possible source of confusion regarding the dependence on auxiliary
structures is the following. One might argue, for instance, that if in the quantum
theory one does not have an explicit dependence on the volume $V_o$ of the cell, then
one can just fix any value for this quantity and one does not have to bother to
change this value given that $V_o$ does not appear  in any of the resulting
expressions. This argument is flawed due to the following reason.
First, one should notice that even when the quantities $(c,p)$ might lead to expressions
that do not contain
explicitly the volume of the cell, these quantities were defined as functions of $V_o$,
precisely to make the symplectic structure well defined and independent of $V_o$.
If at the end of the day we went back to the quantities $(\tilde{c},\tilde{p})$ that
are \emph{truly independent of any cell}, then the quantity $V_o$ would reappear all over
the place. An example,
being the expression for energy density at the bounce.
Thus, one can not simply forget that one introduced an auxiliary fiducial structure,
the cell ${\cal V}$, in the intermediate process. This would be similar to in the quantum treatment
of electromagnetic modes in which one introduced a finite box and then simply forgets to
take the limit
for the box to grow to an infinite size, and then argue that \emph{any} infra red regulator is
physically viable and equally possible. Thus, for the case of loop quantization, the
argument described above misses one basic point regarding the use of auxiliary constructs,
namely that they have to be taken as useful in intermediate steps only,  for which any change
in their allowed values has to be considered, and that the physics should be independent of
any of those choices.

\vskip0.5cm

To summarize this section, we have seen that though it may be straight forward to try to
define a set of phase space variables for $k$=0 isotropic and homogeneous model and quantize
the theory by using exponentiated observables (motivated by holonomies), it is not  true that
one would obtain a sensible quantum theory which has a physically meaningful predictions and a
well defined classical limit. It is very important to understand the
role of the auxiliary structures. It is remarkable that the process of loop quantization picks up
the correct choice of invariant $\beta$ naturally, distinguishing its physical viability from the severe limitations of other inequivalent quantizations.  

Let us end this section with a remark. It may happen that one does not need to introduce
an auxiliary structure such as a cell if the spatial topology is compact (as happens in the $k$=1 FRW or if one considers a flat $k$=0 model on a torus). As we have seen when analyzing the critical density and as we
shall see in the following section, there are more physically motivated conditions
that need to be satisfied by any viable physical theory. In particular, apart from a well defined Planck scale, a `low curvature limit' should also exist. These conditions turn out
to be sufficient to rule out some quantizations.

\section{Departures from General Relativity and various parameterizations}
\label{sec:4}

We now elaborate the way various parameterizations lead to different dynamics from GR in a
large universe depending on the choice of matter. This not only helps
in classifying the energy conditions under which various parameterizations fail, but this
also makes useful to search for a consistent theory for the cases where we have a compact
universe such as a torus topology, where the need of introducing an auxiliary structure does
not arise.
The arguments presented here are also of a slightly different nature than those presented
earlier, since
they are not based on a detailed quantization but rather on the effective dynamics that is
expected to  arise.

We  analyze the dynamics in a flat, $k$=0 and isotropic universe sourced with a single
component of matter, for which we assume for simplicity a fixed equation of state  
$w := P/\rho$ where $P$ is the pressure.
For the Friedman dynamics, the integration of the conservation equation
\be
\dot \rho + 3 H \, (\rho + P) = 0
\ee
leads to
\be
\rho = \rho_o \, \left(\f{a}{a_o}\right)^{-3(1 + w)} ~
\ee
where $\rho_o$ and $a_o$ are constants of integration.
On using Friedman equation we obtain
\be
\dot a  \, \propto \, \rho^{1/2} \,  a \, \propto \, a^{-(3 w + 1)/2} ~.
\ee
Further, the spacetime curvature which is measured by the Ricci scalar:
$R = 6 \left(H^2 + \f{\ddot a}{a} \right)$
can be
computed using Raichaudhuri equation
\be\label{rai}
\f{\ddot a}{a} = - \f{4 \pi G}{3} \, \rho \, \left(1 - 4 \f{\rho}{\rcr} \right) - 4 \pi G \, P \, \left(1 - 2  \f{\rho}{\rcr} \right) ~
\ee
as
\be
\label{Ricci}
R = 6 \left(H^2 + \f{\ddot a}{a} \right) =  8 \pi G \rho \, \left(1 - 3 w + 2 \f{\rho}{\rcr} (1 + 3 w) \right) ~.
\ee
Thus it scales
the same way as $\rho$ i.e. $R \propto a^{-3(1 + w)}$.

We are now equipped to answer the question of when departures from GR occur,
given a particular parameterization used for loop quantization. (Without any loss of
generality, we will restrict our
discussion to the case of expanding universe).
These departures become manifest when the field strength operator in the quantum constraint
differs significantly from the classical analog, which occurs when the exponentiated
quantities (or holonomies) are near saturation.
Thus, a necessary condition for the polymer quantization to yield GR (+ matter) as a
classical limit at low curvatures is that the
phase space variable which is exponentiated must not increase unboundedly as the universe
expands (for those cases in which the large universe limit also corresponds to low
spacetime curvature).

As our first example, let us consider the case of $(P_a, a)$ Wheeler-DeWitt variables. The variation of
$P_a$ with scale factor is%
\be
P_a \, \propto \, \dot a \, a \, \propto \, a^{(-3 w + 1)/2} ~.
\ee
Thus $P_a$ increases as the universe expands when ever $w < 1/3$ which includes various forms of
matter, such as
dust $(w = 0)$,
cosmic strings $(w = -1/3)$ or a cosmological constant $(w = -1)$. As an example of significant
deviations from GR at large scales,
in this quantization once the universe enters a dust dominated epoch, the theory would
predict  unphysical ``quantum gravity effects'' at small curvatures. Such effects would include
a recollapse of the
universe in the classical epoch. This phenomena is inevitable unless the universe exits from the
epoch with $w < 1/3$
to one with $w > 1/3$ at an appropriate time. Since sources with the latter equation of state
decay faster than the
ones with former equation of state in an expanding universe, in any realistic cosmological
scenario this
quantization faces severe problems.

We discussed before that in the old quantization of LQC, it was noted that the universe
recollapses at low curvatures when dominated by a cosmological constant. It is now easy to
understand that this is
bound to happen not only for the case of cosmological constant but when ever dynamics is
dominated by matter which violates strong energy condition. In the old quantization the phase
space variable
which is exponentiated is $c$ which scales as %(taking $N=1$)
\be
c  \, %= \, %\gamma V_o^{1/3} \f{\dot a}{N} \,
\propto \, a^{-(3 w + 1)/2} ~.
\ee
Hence, $c$ increases in an expanding universe for $w < -1 /3$. Thus when matter violates
strong energy condition, the  loop quantized universe would show gross departures from GR at
low curvatures and would recollapse.\footnote{Though a very fine tuned set of initial conditions
may avoid recollapse for some time interval, it is inevitable if the epoch lasts long enough.
Such a problem of fine tuning is over and above the problems associated with the choice of
fiducial cell discussed in the previous section.} Such a quantization will face severe
problems to have any viable
inflationary or dark energy dominated period.

Let us now consider the case of loop quantization which is based on $(\b,\nu)$ variables.
Since $\b = c/|p|^{1/2}$, in the regime where the universe is classical
$\b \propto a^{-3(1 + w)}$. This implies that $\b$ decreases with the expansion of the universe
whenever $w \geq -1$ and increases when ever $w < -1$, in agreement with the behavior of the
spacetime curvature.
Hence {\it for all matter which obeys the null energy condition, LQC leads to an agreement with classical GR at
low curvatures.} This is confirmed by numerical simulations
with massless scalar ($w = 1$) and the cosmological constant ($w = -1$) \cite{aps2}.
Departures from classical GR are however expected for matter which violates null energy condition
such as a phantom field which has $w < -1$ \cite{phantom1}. Interestingly, for such matter content,
the space-time curvature increases as the universe expands, eventually leading to a big rip
singularity.
In this case effective dynamics of LQC predicts a recollapse of the universe which will cure
the big rip singularity \cite{phantom2}.

For the general variable, $P_g$ the variation is given by
\be
P_g \, = c \, p^m \, \propto \, \dot a \, a^{2 m} \, \propto \, a^{-(3 w + 1 - 4 m)/2} ~.
\ee
Thus $P_g$ decreases with expansion of the universe if and only if
$w \geq (4 m - 1)/3$. For $-1/2 < m < 0$, $P_g$ increases with expansion of the universe when $-1 < w < -1/3$ with
a simultaneous decrease of the space-time curvature.
Thus, for this choice of parameters the
evolution leads to significant departures from classical GR at low curvature scales. All such
values of $m$ are thus problematic to yield viable inflationary and dark energy phase in the universe.
Interestingly, if we allow  $m > 0$ (which does not fall in the lattice refined model of Ref.
\cite{lattice}), departures from classical GR can be witnessed without invoking matter which
violates strong energy condition as for example in the case of quantization based on
$(P_a,a)$. Similarly, for $m < -1/2$ departures from GR would not be visible even for $w < -1$. A phantom model
based on such a choice would suffer from the problem of big rip singularity.

To summarize and to make contact with the `lattice refinement approach', we have found that,
by looking at the relation between the variable $P_g$ that becomes `polymerized' and the scale
factor $a$ as given by the Friedman dynamics (assumed to be valid at some regime), we have
analyzed for several equations of state the behavior of the polymer approximation.
Within the range of the parameter $m\in(-1/2,0)$ allowed within the `lattice refinement'
approaches \cite{lattice}, we have seen that there will always be spurious quantum effects
coming from the loop quantization.
This leads us to conclude
that these models can safely be considered, at best, phenomenologically inviable. Further, {\it if we demand that
the quantum theory should approximate GR at low curvatures for all matter satisfying null energy condition then
all the inequivalent quantizations to LQC are ruled out.}

\section{Discussion}
\label{sec:7}

Let us summarize our results. We have analyzed the physical and phenomenological implications
of loop quantized $k$=0 FRW models, for a one parameter family of inequivalent quantization
prescriptions (including the original LQC model and the so called `lattice refining models'). This ambiguity in the
quantization can be recast in terms of the choice of basic phase space variables, that serve
as starting point for the loop quantization.
In order to analyze these models, we considered the role of auxiliary structures --such as the fiducial cell needed for this case-- and
focused our attention on two fronts, namely the high energy density/`Planck scale' as
defined by each of these models, as well as on their `low energy', classical limit. 
We have shown that, for the family considered, there is a unique prescription for which the resulting quantum theory is
independent from the choice of fiducial cell and is physically viable both at Planck and low
curvature scales. This quantization corresponds precisely to the improved dynamics \cite{aps2,slqc}, the one which uses the coordinates $(\b,\nu)$, associated to the physical volume of a fiducial cell and its conjugate variable.

A conclusion one may draw from these considerations is that in order to find a
consistent quantization, when addressing more complicated models, one can use the criteria advocated in this paper as a starting point in the
quantization process. 
For instance, in recent literature attempts have been
made to study the viability of lattice refinement models through stability analysis of difference
equations \cite{mb_cartin_khanna} and using phenomenological approaches \cite{lattice_works}.
However, above
consistency requirements  remain unconsidered,  especially the role of auxiliary structures.
Given that self-consistent quantizations are limited,  it is  not surprising that some of these
works point to
problems and conclusions for lattice models similar to those reached here in the present analysis.
In our view, before exploring the details of stability
and phenomenological consequences of arbitrary quantizations, even if one requires that the considered
model passes
consistency requirement regarding fiducial structures, many valuable lessons can be learnt at an early stage and various models
can be ruled out  at a preliminary level\footnote{One can also demand these consistency conditions
to be
simultaneously satisfied in the stability analysis and phenomenological studies. For example,
in the numerical analysis of
properties of difference equations for general refinements one should not suppress the dependence on the fiducial cell.
It is then expected  that stability methods will lead to stronger conclusions
\cite{vh_private}.}.

An important lesson one might draw from this simple model is that it is indeed possible to guide the loop quantization process by means of both self-consistency and physical viability. It is not true that `anything goes' in loop quantization. Given the limitations of an arbitrary uniform discretization in geometric
variables such as scale factor, it may be asked whether their suitable
refinement  would work. From our results one may conclude that if a refinement
is made such that the uniform discretization appears as in the improved dynamics of LQC, one will obtain sensible results. Then the descriptions will be equivalent.
However, such a refinement will
be ad-hoc unless it is separately justified in the model. Note that at the
classical level there is no motivation to consider such a discretization and
its refinement. This has to be contrasted with LQC where the difference equation equally spaced in volume is forced by the underlying quantization procedure.

The issue of ambiguities that we have addressed in the isotropic models can also be
similarly tackled in the anisotropic and black hole interior models in
loop quantization. The status of the quantum theory and a physically viable description in both the cases is still in its early stages. However, motivated by the improved
quantization of LQC, effective models have been constructed \cite{bianchi,b-k}.
A straightforward analysis, along the lines here presented, shows that in
these models one faces the challenge of
having a consistent description in terms of a uniform discretization in terms of
some phase space variables.
This signals that the direct application of methods of the isotropic sector
proves insufficient and may very well be misleading. For example, so far in Bianchi models the role of energy density could be over emphasized. In isotropic LQC, with a fixed equation of state, it directly measures the spacetime curvature and hence can be associated an invariant meaning. This fails to be true in the anisotropic models. Thus, it is
pertinent to ask questions directly about invariant entities like Ricci curvature instead of concluding about the nature of bounce from energy densities or volumes in more general models. It is not surprising that
some of these constructions lead to unnatural physical effects such as Planck
scale phenomena near coordinate singularities \cite{b-k}.
Thus, it is important to apply the guiding principles as we have investigated
here in isotropic models to such constructions to understand the mathematical and
physical consistency.

Having said this, one must remark, however, that one should not expect a
universal recipe for approaching a fully successful loop quantization.
We have put forward some consistency and
physically motivated conditions that any quantum theory must satisfy.
Even when they have proved to be highly successful for the case of isotropic cosmologies,
one can not expect them to provide a  royal road for the loop quantization prescription in general.
These conditions can be regarded as necessary for a consistent quantization, but they are
by no means sufficient, since their particular implementation can depend on the details
of the system under study. For instance, one expects that for less symmetric models, such as Bianchi cosmologies and the Schwarzschild interior, even when guided by these criteria to select the --still to be constructed-- physically viable quantization(s), the particular implementation might require some adjustments.
For instance, as we have mentioned above, the criteria for
specifying what the right `Planck scale' is, that in the isotropic sector is set by the energy density, could be replaced by more `covariant' curvature invariants.

As usual, due care must be exercised in each case which brings up new challenges.
However, one can be optimistic that a systematic study of  symmetric systems will
bring us closer to the main goal of defining a physically relevant quantization in full LQG.

\section*{Acknowledgments}

\noindent
We thank A. Ashtekar, G. Mena-Marugan, T. Pawlowski, S. Speziale and
K. Vandersloot for discussions
and comments.
This work was in part supported by CONACyT U47857-F
grant, by NSF PHY04-56913 and by the Eberly Research Funds of Penn
State. The research of PS is supported by Perimeter Institute for Theoretical
Physics.  Research at Perimeter Institute is supported by the Government
of Canada through Industry Canada and by the Province of Ontario through
the Ministry of Research \& Innovation.

\end{document}